\documentclass[aps,a4paper,prl,twocolumn,superscriptaddress,groupedaddress,11pt]{revtex4-2}  
\usepackage{graphicx}  
\usepackage{dcolumn}   
\usepackage{bm}        
\usepackage{amssymb}   
 \usepackage{float}
\usepackage{amsfonts}
\usepackage{gensymb}
\usepackage{amsmath}
\usepackage{upgreek}
 \usepackage{hyperref}

\hypersetup{
   colorlinks=true,
    linkcolor=blue,
    filecolor=magenta,      
    urlcolor=blue,
}
 
\usepackage{blindtext,tikz}
\usetikzlibrary{calc}
\begin{document}
\title{Evolution of the Chern Gap in Chern Magnet HoMn$_6$Sn$_{6-x}$Ge$_x$}

\author{Christopher Sims}
\affiliation{\textit{ Department of Physics, University of Central Florida, Orlando, Florida 32816, USA}}
\date{\today}
\begin{abstract}

{The Chern gap is a unique topological feature that can host non-abelian particles. The Kagome lattice forms a chern insulator with no gap. Upon the inclusion of magnetism the Kagome system hosts a Chern gap at the K points in the lattice. In this work, the effect of Ge doping on HoMn$_6$Sn$_{6}$ is investigated. It is seen that with increased doping, a multi-stack Chern gap in formed in HoMn$_6$Sn$_{6-x}$Ge$_x$. In addition, the Chern gaps are much more pronounced and larger in energy in HoMn$_6$Ge$_{6}$ then HoMn$_6$Sn$_{6}$.}
\end{abstract}
\date{\today}
\maketitle

\section{Introduction}

Topological quantum materials is a rapidly growing field which investigates novel non-trivial topological states in condensed matter. The discovery of 3D topological insulators with an insulating bulk and a conducting surface with a Dirac cone was the stepping stone which changed the field of condensed matter\cite{Hasan2010,Bansil2016}. Since the discovery of a topological insulator, Weyl and nodal line Fermions have been discovered in TaAs\cite{Xu2015,Lv2015,Huang2015,Weng2015,Nielsen1983,Wan2011} type materials and ZrSiS type materials\cite{Neupane2016a,Hosen2017,Hu2016a,Takane2016,Hu2017,Hosen2018a,Hosen2018b}, respectively. Furthermore, Dirac and Weyl semimetals have been discovered to violate Lorentz invariance in their overtilted type-II phase in materials such as type-II Weyl semi-metal CeAlSi or type-II Dirac semimetal PtTe$_2$. These discoveries have been focused on the topological invariant $\mathbb{Z}_2$. There exists another topological invariant that also hosts topologically nontrivial states, the Chern invariant $\mathbb{C}$.

In his seminal paper\cite{Haldane1988}, Haldane formulated a tight binding model that was able to predict the formation of topologically non trivial states in the hexagonal lattice, this was latter discovered in cold atom systems\cite{Jotzu2014,Skirlo2015} and in condensed matter as Dirac Fermions \cite{Thonhauser2006}. The Haldane tight binding model was latter extended to the Kagome lattice where there is a closing of the band gap at the K points and the formation of a Chern insulator \cite{Guo2009,Ghimire2020}. Further investigation into the Kagome lattice showed that the anomalous hall effect was due to the Chern number of the system. Recently, Kagome systems\cite{Kang2019} have been discovered to host large anomalous hall effects in Mn$_3$X (X = Ge, Sn)\cite{Nayak2016,Yang2017}.

Recently, the HfFe$_6$Ge$_6$ type Kagome system \cite{Idrissi1991,Venturini1993,Zhang2001,Shaoying2001,Mazet2006,Mazet2007} has been rediscovered as an ideal system to study the Chern insulating state. Interestingly, the onset of magnetism in these materials show the formation of the Chern gap, both in theoretical predictions and in experimental measurements \cite{Yin2020,Ghimire2020a,Chen2021}. The  Kagome system crystallizes in space group 191 (P6/mmm) and has over 200 unique stoichiometric materials making this system one of the most tunable condensed matter systems explored in recent times. Although this system is highly tunable, there is a large interest in RMn$_6$Sn$_6$ (R = rare earth elements)\cite{Ma2021}, which can be easily grown with the tin flux technique. The Kagome system has one R atom in the center with triangles at the K points composed of magnetic Mn atoms and nonmagnetic Sn atoms.

There is an increasing interest in magnetic Kagome materials that host the Chern gap due its unique feature. Chern gaped Dirac cones host topologically protected non-abelian anyons. non-abelian anyons, similar to the Majorana Fermion, allow for topologically protected quantum computing \cite{Sarma2005,Nayak2008,Stern2013}. Recently Anyons have been directly measured in fractional Hall effect interferometers \cite{Halperin2011,Nakamura2020,Bartolomei2020}. However, these Anyons obey abelian statistics which makes them topological trivial. By tuning the parameters in the extensive Kagome system researchers aim to discover a wide Chern gap material that hosts non-abelian physics.

\begin{figure*}[ht]
 \centering
 \includegraphics[width=1\textwidth]{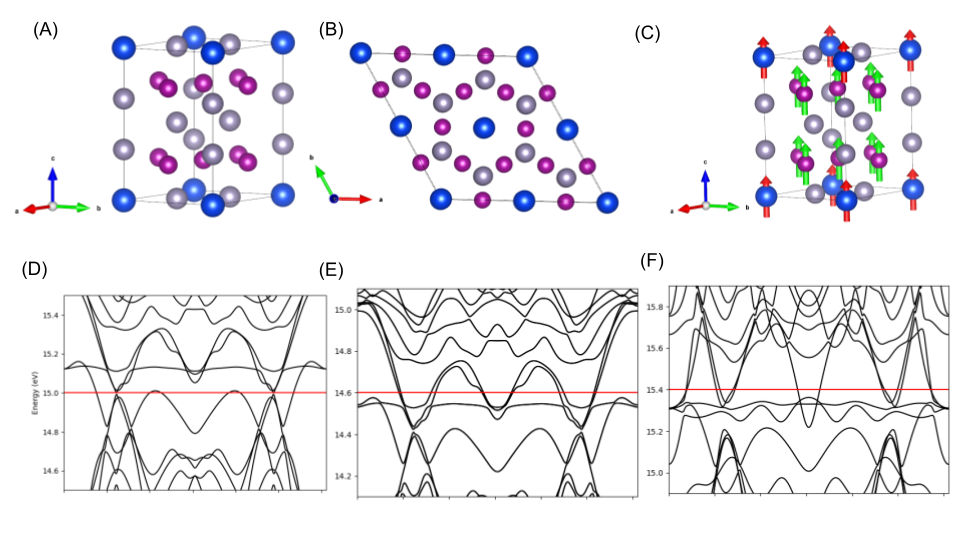}
  	\caption{\textbf{Bulk Band Structure of HoMn$_6$Sn$_{6}$:} (A) isometric view of HoMn$_6$Sn$_{6}$ crystal structure (B) (001) view of the HoMn$_6$Sn$_{6}$ crystal (C) View of the magnetic moments of HoMn$_6$Sn$_{6}$ (D)
bulk band struture in the $\Gamma$-K-M-$\Gamma$ direction (D) without SOC (D) with SOC (F) with SOC and magnetism}
\label{BULK}
 \end{figure*}

\indent This work provides provides a theoretical investigation into the effect of changing one of the parameters in this highly tunable system, and the effect it has on the Chern gap formation. The Tin (Sn) atoms in HoMn$_6$Sn$_{6}$ is gradually doped with Germanium (Ge)  HoMn$_6$Sn$_{6-x}$Ge$_x$ (x = 0, 2, 4, 6) in order to see the effect on the bands structure. Interestingly the Chern gap disappears with light doping and then forms two new stacked chern gaps in HoMn$_6$Ge$_6$. This multigap feature has been previously observed in the Chiral semimetal Rh$_{0.955}$Ni$_{0.045}$Si\cite{Cochran2020} which supports the finding that multi-stack Chern gaps exist.

\begin{figure*}[ht]
 \centering
 \includegraphics[width=0.8\textwidth]{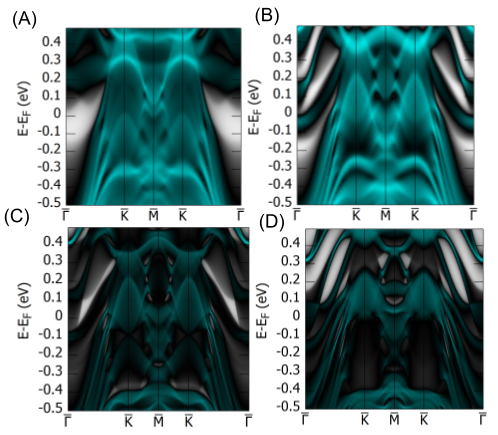}
  	\caption{\textbf{Surface States of HoMn$_6$Sn$_{6-x}$Ge$_x$:} (A) HoMn$_6$Sn$_{6}$ (B) HoMn$_6$Sn$_4$Ge$_2$ (C) HoMn$_6$Sn$_2$Ge$_4$ (D) HoMn$_6$Ge$_6$}
\label{SS}
  \end{figure*}

\section{Methods}
HoMn$_6$Sn$_{6-x}$Ge$_x$ crystallizes in the Kagome HfFe$_6$Ge$_6$-type structure (space group No. 191, P6/mmm) with lattice parameters a = b = 5.061 \AA, and c = 8.083 \AA.

The band structure calculations were carried out using the density functional theory (DFT) program Quantum Espresso (QE)\cite{wannier90}, with the generalized gradient approximation (GGA)\cite{pbegga} as the exchange correlation functional. Projector augmented wave (PAW) pseudo-potentials were generated utilizing PSlibrary. The relaxed crystal structure was obtained from materials project (mp-19725)\cite{Jain2013} via the CIF2WAN interface \cite{Sims2020a}. The energy cutoff was set to 100 Ry and the charge density cutoff was set to 700 Ry for the plane wave basis with a k-mesh of 25 $\times $25 $\times $25.  After the self consistent calculation was completed (SCF), the Wannier tight binding Hamiltonian was generated from the non-self consistent calculation (NSCF) with Wannier90\cite{wannier90}. The surface spectrum \cite{sancho1985} was calculated with Wannier Tools \cite{wtools}. For HoMn$_6$Sn$_{6}$ these calculations were performed without spin orbit coupling (SOC), with spin-orbit coupling, and with SOC and magnetism to confirm the robustness of the calculations. All other doping values were calculated with SOC and magnetism. The final magnetism is found to be 0.18 $\mu_B$ for Ho and and 2.34 $\mu_B$ for Mn for HoMn$_6$Sn$_{6}$. These magnetic values vary only slightly with different doping values.

\begin{figure*}[ht]
 \centering
 \includegraphics[width=0.8\textwidth]{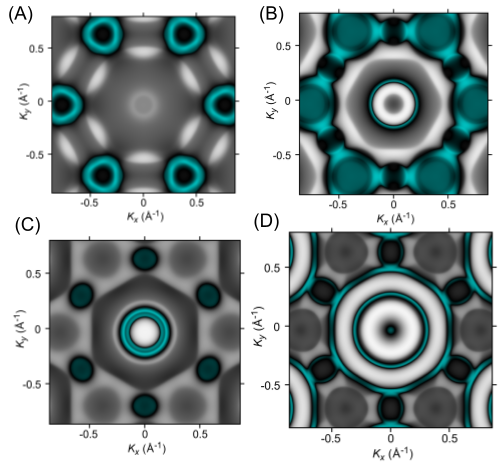}
  	\caption{\textbf{Chern Gap States on the (001) surface}. (A) HoMn$_6$Sn$_{6}$ (B) HoMn$_6$Sn$_4$Ge$_2$ (C) HoMn$_6$Sn$_2$Ge$_4$ (D) HoMn$_6$Ge$_6$}
\label{ARC}
 \end{figure*}

\section{Results and Discussion}
The crystal structure of HoMn$_6$Sn$_{6}$ is presented in the isometric direction [Figure \ref{BULK}(A)], HoMn$_6$Sn$_{6}$ is composed of stacked layers of HoSn$_2$ and SnMn$_3$, with an intermediate Sn$_2$ layer. By viewing the (001) surface of a 2x2 lattice [Figure \ref{BULK}(B)], the Kagome structure can be seen where there is a hexagonal ring composed of Mn-Sn atoms around the central Ho atom at the conventional cell's center and triangles with a central Sn atom surrounded by 3 Mn atoms at the cell's corners. When the system becomes magnetic, Ho and Mn magnetize to the C-axis in a ferromagnetic configuration [Figure \ref{BULK}(C)]. A band structure diagram is presented without spin orbit coupling (SOC) [Figure \ref{BULK}(D)] in the $\Gamma$-K-M-K-$\Gamma$ direction, a clear gap is seen at the K points. With the inclusion of spin orbit coupling (SOC) [Figure \ref{BULK}(E)] the gap still exists in the bulk, however the surface band calculation show that there is a Dirac cone inside the gap. When magnetism is added, the gap widens significantly at the K point to be about 0.3 eV. In addition, the Dirac cone becomes gaped and forms a Chern gaped Dirac cone, which is confirmed by Wilson loop analysis.

\begin{figure*}[!ht]
 \centering
 \includegraphics[width=0.8\textwidth]{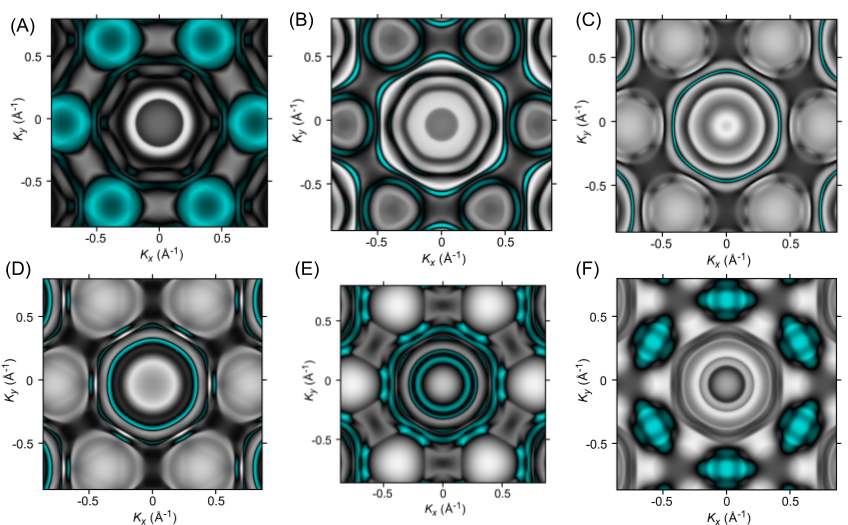}
  	\caption{\textbf{Energy Contours of HoMn$_6$Ge$_{6}$ }. (A) E-E$_F$= 100 meV (B) E-E$_F$= 0 meV (C) E-E$_F$= -100 meV (D) E-E$_F$= -200 meV (E) E-E$_F$= -300 meV (F) E-E$_F$= -400 meV}
\label{EC}
 \end{figure*}

Figure \ref{SS} shows the evolution of the surface electronic structure in the $\mathrm{\overline{\Gamma}}$-$\mathrm{\overline{K}}$-$\mathrm{\overline{M}}$-$\mathrm{\overline{K}}$-$\mathrm{\overline{\Gamma}}$ direction with different doping levels of Ge. Pure HoMn$_6$Sn$_{6}$ [Figure \ref{SS}(A)] shows a Chern invariant about 0.3 eV above the Fermi level, where the Chern gap forms. The Chern gap can be seen in Figure \ref{SS}(A) by examining the two intense bands that arise from the hybridization at about 0.4 eV and 0.25 eV. As doping increases, the Chern invariant goes away above the Fermi level in HoMn$_6$Sn$_{4}$Ge$_2$ [Figure \ref{SS}(B)], this is likely because the two bands no longer interact. The topological invariant reappears with higher doping concentrations HoMn$_6$Sn$_{2}$Ge$_4$ [Figure \ref{SS}(C)], and HoMn$_6$Ge$_6$ [Figure \ref{SS}(D)] from further hybridization of the bands. Interestingly, it can be seen that with increasing Ge doping the electron bands at the $\mathrm{\overline{K}}$ point continue to go lower in energy and the hole bands at the $\mathrm{\overline{K}}$ point go up in energy. These bands hybridize to form the original Chern gap in HoMn$_6$Sn$_{6}$, however in HoMn$_6$Ge$_6$ these bands now hybridize to form two Chern gaps at the $\mathrm{\overline{K}}$ point. These stacked, ladder-like Chern gaps are larger and more distinct then those in HoMn$_6$Sn$_{6}$ and should allow for more prominent Chern gap features to be noticed, such as the anomalous Hall effect (AHE). In figure \ref{SS}(C) (HoMn$_6$Sn$_{2}$Ge$_4$) there is a critical cross over where the bands begin to form a more distinct Chern gap at $E-E_F$ = 0.2 eV, -0.1 eV, -0.2 eV. The stacked Chern Gaps in HoMn$_6$Ge$_6$ have a dispersion of 0.2 eV and 0.4 eV respectively.

In order to elucidate the nature of the DOS in the Chern gap, the (001) surface states are calculated in the Chern Gap [Figure \ref{ARC}]. HoMn$_6$Sn$_{6}$ shows low intensity in the Chern gap except triangle like features around the $\mathrm{\overline{K}}$ points [Figure \ref{ARC}(A)], the low intensity at the center of the triangle features is the Chern gap state. The bands in the Chern gap are calculated to be topologically non trivial. The trivial states in HoMn$_6$Sn$_{4}$Ge$_2$ [Figure \ref{ARC}(B)] shows bulk bands that persist at the $\mathrm{\overline{K}}$ points. In the center a hexagonal, low DOS, feature is seen. HoMn$_6$Sn$_{2}$Ge$_4$ shows a clear gap with low intensity from electron-like bands that form the secondary Chern gap lower in binding energy [Figure \ref{ARC}(C)]. It is noted that the hexagonal like feature seen in [Figure \ref{ARC}(B)] is rotated by 15$^{\circ}$, this shows a clear change in the nature of the bands that cross the Chern gap in the doping regime between X=2 and X=4. HoMn$_6$Ge$_6$  [Figure \ref{ARC}(D)] shows a clear Chern gap with low intensity electron bands that also cross the gap, these bands form the secondary Chern gap lower in binding energy as seen in Figure \ref{SS}(D).

The multi-stack in HoMn$_6$Ge$_6$ is an interesting feature that warrants further theoretical investigation. Constant energy contours are calculated for E-E$_F$ = 100 meV, 0 meV, -100 meV, -200 meV, -300 meV, -400 meV. Slightly above the Fermi level  [Figure \ref{EC}(A)], the $\mathrm{\overline{K}}$ points posses the bulk bands that separate the main Chern gap above the Fermi level [Figure \ref{ARC}(D)] and the secondary Chern gap below the Fermi level. At the Fermi level [Figure \ref{EC}(B)], there are 4 bands that enclose the $\mathrm{\overline{\Gamma}}$ point, from the center they are circular, hexagonal, hexagonal, and a hexagon rotated by 15$^{\circ}$. These bands cross the Chern gap and form the main Chern gaped Dirac cone. 200 meV below the Fermi level the secondary Chern gap begins to form, as opposed to the Chern gap in HoMn$_6$Sn$_{6}$, the $\mathrm{\overline{K}}$ points do not have bands that interfere with the Chern features. As the binding energy increases the Chern gap stays persistent and the $\mathrm{\overline{\Gamma}}$ centered hexagonal bands continue to shrink [Figure \ref{EC}(D-E)] until they end around 400 meV below the Fermi level [Figure \ref{EC}(F)].

\section{Conclusion}
In conclusion, with increased Ge doping, the Chern gap in HoMn$_6$Sn$_{6}$ closes and then forms a multi-stack Chern gap in HoMn$_6$Ge$_{6}$. This theoretical investigation provides an intrigue to study topological Chern gaps in RMn$_6$Ge$_{6}$ which could potentially host novel topological effects due to the existence of two Chern gaps in this material. This discovery promotes further investigation with scanning tunneling microscopy (STM), Angle resolved photo emission spectroscopy (ARPES), and electronic measurements.

 \section{Acknowledgments}
 \vspace{-4mm}
 The authors acknowledge the University of Central Florida Advanced Research Computing Center for providing computational resources and support that have contributed to results reported herein. URL:\href{https://arcc.ist.ucf.edu}{https://arcc.ist.ucf.edu}.
 
 Correspondence and requests for materials should be addressed to C.S.
 (Email: Christophersims@knights.ucf.edu)

 \section{Supplementary}
  \vspace{-4mm}
See supplementary for further details on DFT calculations, topological invariant calculations and the surface band structure of HoMn$_6$Sn$_{6}$ with and without SOC.

%

\end{document}